# Relevant Aperiodic Modulation in the $2d$ Ising Model.

F. IGLÓI[1,2] and L. TURBAN[1]

[1] *Laboratoire de Physique du Solide*(*), *Université de Nancy I*
*BP 239, F-54506 Vandœuvre lès Nancy Cedex, France*
[2] *Reasearch Institute for Solid State Physics*
*P.O. Box 49, H-1525 Budapest 114, Hungary*(**)

**Abstract.** – We consider the surface critical behaviour of a semi-infinite two-dimensional layered Ising model, where the couplings perpendicular to the surface follow the aperiodic Rudin-Shapiro sequence. The model has unusual critical properties: depending on the strength of the modulation, at the bulk critical point the surface magnetization is either discontinuous or vanishes with an essential singularity. The critical surface magnetization as well as the form of the essential singularity are calculated exactly.

*Introduction.* – Motivated by the discovery of quasi-crystals [1] and the possibility to build artificial layered structures (see for example [2]), the theoretical study of phase transitions and critical phenomena in quasi-periodic or, more generally, aperiodic systems has considerably developed in the last years (for a review, see [3]).

Following a series of numerical [4-9] and analytical [10-18] studies, an extension to aperiodic perturbations of the Harris criterion [19] for disordered systems was recently proposed [20]. According to the Luck criterion, an aperiodic modulation can be irrelevant, marginal or relevant, depending on the value of a crossover exponent involving the correlation length exponent $\nu$ and the wandering exponent $\omega$, characteristic of the sequence. The form of the singularities for relevant modulations and the case of anisotropic critical systems were discussed in [21] using scaling arguments. The criterion, first established for uniaxial aperiodicities, was extended to higher-dimensional ones in [22].

In this work we present exact results for the case of a relevant aperiodic modulation of the interactions following the Rudin-Shapiro (RS) sequence [23, 20] in the layered $2d$ Ising model. Thus we complete a previous study [24] where exact results were obtained for irrelevant

---

(*) Unité de Recherche Associée au CNRS No. 155.
(**) Permanent address.

Typeset using EURO-LATEX

and marginal perturbations, whereas the relevant case, which is the most difficult to handle analytically, was only treated numerically and through scaling considerations.

We consider the Ising model on a semi-infinite square lattice in which the couplings parallel to the free surface are constant and the modulation is associated with the perpendicular couplings. The problem is treated in the extreme anisotropic limit [25] where the transfer operator along the surface involves the Hamiltonion of a spin-$\frac{1}{2}$ Ising quantum chain

$$\mathcal{H} = -\frac{1}{2}\sum_{k=1}^{\infty}[\lambda_k \sigma_k^x \sigma_{k+1}^x + \sigma_k^z]. \tag{1}$$

The surface magnetization can be obtained using standard methods [26, 27] as a function of the aperiodic couplings $\lambda_k$ with

$$m_\mathrm{s} = \left(1 + \sum_{j=1}^{\infty}\prod_{k=1}^{j}\lambda_k^{-2}\right)^{-1/2}. \tag{2}$$

Let us associate with each bond a digit $f_k$ which can take two values, 0 or 1, depending on the type of the bond along the aperiodic sequence and let $\lambda_k = \lambda r^{f_k}$. The surface magnetization in eq. (2) can then be rewritten as

$$m_\mathrm{s} = \left[S\left(\frac{1}{\lambda^2 r}, r\right)\right]^{-1/2}, \quad S(x,r) = \sum_{j=0}^{\infty} x^j r^{j-2n_j}, \quad n_j = \sum_{k=1}^{j} f_k, \quad n_0 = 0. \tag{3}$$

The critical coupling $\lambda_\mathrm{c}$ is such that $\lim_{j\to\infty}(1/j)\sum_{k=1}^{j}\ln\lambda_k = 0$ [28], so that one obtains $\lambda_\mathrm{c} = r^{-\rho_\infty}$, where $\rho_\infty$ is the density of 1 in the infinite sequence.

*Substitution matrix and scaling considerations.* – An aperiodic sequence can be constructed through iterated substitions on the letters $A, B, \ldots$. The RS sequence itself follows from

$$A \to \mathcal{S}(A) = AB, \quad B \to \mathcal{S}(B) = AC, \quad C \to \mathcal{S}(C) = DB, \quad D \to \mathcal{S}(D) = DC. \tag{4}$$

Informations about the properties of the sequence are contained in its substitution matrix [29] with entries $n_i^{\mathcal{S}(j)}$ giving the number of letters $i$ in $\mathcal{S}(j)$ ($i,j = A, B, \ldots$). The length of the sequence after $n$ iterations is related to the largest eigenvalue $\Lambda_1$ through $L \sim \Lambda_1^n$ and the letter densities to the corresponding eigenvector.

With the $k$-th letter in the sequence, one associates a coupling which is either $\lambda$ or $\lambda r$, according to the value of $f_k$ (0 or 1). The cumulated deviation $\Delta(L)$ from the average coupling $\overline{\lambda}$ at the length scale reached after $n$ iterations can be expressed as

$$\Delta(L) = \sum_{k=1}^{L}(\lambda_k - \overline{\lambda}) = \lambda(r-1)(n_L - L\rho_\infty) \sim \delta|\Lambda_2|^n \sim \delta L^\omega, \tag{5}$$

where $\delta \sim r - 1$ and the wandering exponent $\omega$ [29] is given by $\ln|\Lambda_2|/\ln\Lambda_1$, where $\Lambda_2$ is the next-to-leading eigenvalue. Looking at the behaviour of the thermal perturbation $\Delta(L)/L \sim \delta L^{\omega-1}$ under a change of scale by $b = L/L'$, one obtains the scaling behaviour for the amplitude [20, 21]

$$\delta' = b^{\Phi/\nu}\delta, \qquad \Phi = 1 + \nu(\omega - 1). \tag{6}$$

The relevance of the perturbation depends on the sign of the crossover exponent $\Phi$. For the $2d$ Ising model with $\nu = 1$, the aperiodic modulation is irrelevant when $\omega < 0$, marginal when $\omega = 0$ and relevant when $\omega > 0$. In the marginal case, varying exponents are obtained [21, 24].

The behaviour of the system in the vicinity of the critical point follows from scaling considerations. The surface magnetization behaves as [24]

$$m_{\rm s}(t,\delta) = t^{\beta_{\rm s}} F(u), \qquad u = \frac{l_{\rm a}}{\xi}, \tag{7}$$

where $t = 1 - (\lambda_{\rm c}/\lambda)^2$, $\beta_{\rm s}$ is the magnetization exponent at the ordinary surface transition, $\xi \sim t^{-\nu}$ is the bulk correlation length and the scaling variable $u$ involves a new characteristic length associated with the aperiodicity, $l_{\rm a} \sim |\delta|^{-\nu/\Phi}$, which stays finite at the bulk critical point.

*Surface magnetization for the Rudin-Shapiro sequence.* – For the RS sequence, one may associate $f_k = 0$ with $A$ and $B$ and $f_k = 1$ with $C$ and $D$ in (4), starting the sequence on $A$. The largest eigenvalues in modulus of the substitution matrix are $\Lambda_1 = 2$, $\Lambda_2 = \pm\sqrt{2}$, leading to the wandering exponent $\omega = \frac{1}{2}$. The asymptotic densities of the letters yield $\rho_\infty = \frac{1}{2}$ thus the critical coupling is $\lambda_{\rm c} = r^{-1/2}$.

Using the two-digit correspondence $A, B, C, D \to 00, 01, 10, 11$ which follows from (4) after one iteration, one obtains the following recursion relations for the $n_j$'s:

$$\begin{cases} n_{8p} = 2p + 2n_{2p}, & n_{8p+1} = 2p + 2n_{2p} + f_{2p+1}, \\ n_{8p+2} = 2p + 2n_{2p+1}, & n_{8p+3} = 2p + 2n_{2p} + 3f_{2p+1}, \\ n_{8p+4} = 2p + 1 + 2n_{2p+1}, & n_{8p+5} = 2p + 1 + 2n_{2p+1} + f_{2p+2}, \\ n_{8p+6} = 2p + 1 + 2n_{2p+2}, & n_{8p+7} = 2p + 2 + 2n_{2p+1} + f_{2p+2}. \end{cases} \tag{8}$$

This allows us to rewrite the sum in (3) as

$$S(x,r) = -x^{-2}\left(1 + \frac{x + x^{-1}}{r + r^{-1}}\right) + \left[1 + x^{-2} + \frac{x^{-1}}{2}(r + r^{-1}) + \frac{x + x^{-3}}{r + r^{-1}}\right]S(x^4, r^2) \\ - \frac{x^{-1}}{2}\frac{r^2 + r^{-2} + 2x^4}{r + r^{-1}}S(x^4, -r^2), \tag{9}$$

where $x = (\lambda_{\rm c}/\lambda)^2$. Changing $r$ into $-r$ leads to a second equation giving $S(x^4, r^2)$ and $S(x^4, -r^2)$. At the critical point $x = 1$, with $S(1, r) = 1 + g(r)$, one obtains the following functional equations:

$$g(r^2) = \frac{1}{4}[g(r) + g(-r)], \quad g(-r^2) = \frac{(1-r)^4 g(r) + (1+r)^4 g(-r)}{4(1+r^2)^2}. \tag{10}$$

Now, when $r < 1$ one may look for $g(r)$ under the form of a series expansion in powers of $r$, $g(r) = \sum_{p=0}^{\infty} a_p r^p$, where, according to (10), the coefficients satisfy

$$a_{2p} = 2a_p, \quad a_{4p-1} + a_{4p-3} = 4a_{2p-1}, \quad a_{4p+1} + a_{4p-1} = a_{2p+1} + 2a_{2p} + a_{2p-1}. \tag{11}$$

These recursion relations are solved by $a_p = p$, leading to $g(r) = r/(1-r)^2$ and finally, according to (3), the critical surface magnetization is given by:

$$m_{\rm s,c} = \frac{1-r}{\sqrt{1-r+r^2}}, \qquad (r \leq 1). \tag{12}$$

When $r > 1$, $g(r)$ diverges, *i.e.*, the surface magnetization vanishes at the bulk critical point. To study its behaviour on a finite system with size $L$, one has to terminate $S(1, r)$ in (3) at $j = L$. For $L = 2^{2n-1}$ one finds by inspection that it corresponds to the first $2^n$ terms of the power series for $g(r)$, up to a correction for the last one. Thus, one obtains

$$g(r, L = 2^{2n-1}) = \sum_{p=1}^{2^n} pr^p - 2^{n-1}r^{2^n} = \frac{r}{(r-1)^2}\left[1 + \left(\frac{r^2-1}{2r}2^n - 1\right)r^{2^n}\right]. \quad (13)$$

When $n \to \infty$ the previous result is recovered for $r \leq 1$, whereas the expression diverges for $r > 1$. In the scaling limit, $r - 1 \ll 1$, $L \to \infty$, one obtains the following finite-size behaviour for the critical surface magnetization:

$$m_{s,c}(L) \simeq \left[\frac{r-1}{\sqrt{2L}}\right]^{1/2} \exp\left[-\frac{(r-1)\sqrt{2L}}{2}\right]. \quad (14)$$

Thus, the finite-size dependence of $m_{s,c}$, having a stretched exponential form, is anomalous for $r > 1$. A similar behaviour is obtained below for the temperature dependence.

In order to determine the general behaviour of the surface magnetization when $1 \gg t > 0$ and $|r-1| \ll 1$, we take a continuum approximation for the sum in (3) and use (5) to write $j - n_j/\rho_\infty \simeq Aj^\omega$, so that

$$m_s^{-2} \simeq \int_0^\infty dj \exp(-jt + Bj^\omega), \qquad B = A \ln r \simeq A(r-1). \quad (15)$$

When $B < 0$, expanding $\exp(-jt)$ and integrating term by term, one obtains

$$m_s^{-2} \simeq \frac{|B|^{-1/\omega}}{\omega} \sum_{n=0}^\infty \frac{(-t|B|^{-1/\omega})^n}{n!} \Gamma\left(\frac{n+1}{\omega}\right) = \frac{\Gamma(1/\omega)}{\omega|B|^{1/\omega}}\left[1 - \frac{\Gamma(2/\omega)}{\Gamma(1/\omega)|B|^{1/\omega}}t + \cdots\right]. \quad (16)$$

The surface magnetization approaches its nonvanishing critical value as $m_s(t) - m_{s,c} \sim t^{\beta'_s}$, with a critical exponent $\beta'_s = 1$ for any value of the wandering exponent $\omega$. For the RS sequence with $\omega = \frac{1}{2}$, one obtains

$$m_s \simeq \frac{A(1-r)}{\sqrt{2}}\left[1 + \frac{3t}{A^2(1-r)^2} + \cdots\right], \quad (17)$$

in agreement with the exact result in (12) and with the scaling form in (7) with $\beta_s = \frac{1}{2}$, $\nu = 1$ for the 2d Ising model, and $F(u) \sim u^{-1/2}(1 + au + \cdots)$. According to (7), the critical surface magnetization generally vanishes at $r = 1$ as $|r - 1|^{\beta_s/\Phi}$.

When $B > 0$, the integral in (15) can be evaluated using Laplace's method when $0 < \omega < 1$. A straightforward calculation then gives, up to a power law prefactor,

$$m_s \sim \exp\left[\frac{(\omega-1)\,t^{\omega/(\omega-1)}}{2\omega(\omega B)^{1/(\omega-1)}}\right] \sim \exp\left[-\frac{A^2}{8}\frac{(r-1)^2}{t}\right] \quad (18)$$

where the last expression is the RS result.

*Discussion.* – The surface order (12) at the bulk critical point is linked to a local enhancement of the couplings for $r < 1$. Starting the sequence on $D$ would amount to exchange weak and strong couplings and the surface first-order transition would then occur for $r > 1$. Since $\rho_\infty = \frac{1}{2}$, one does not expect such an asymmetry in the bulk critical behaviour.

The amplitude $A$ in (15), which was considered as a constant in the continuum calculation, is actually an oscillating function of $j$ [29]. This does not affect the stretched exponential in (18) but may influence the prefactor which was ignored. In order to determine it, one may use the exact finite-size result (14), replacing $L$ by the relevant thermal length $\tilde{L}$, associated with the aperiodicity, which is obtained by equating the two terms in the exponential of eq. (15). Since $\tilde{L} \sim [(r-1)/t]^{1/(1-\omega)} = [(r-1)/t]^2$ for the RS sequence, the prefactor is proportional to $t^{1/2}$ and the scaling function in eq. (7) is simply $F(u) \sim \exp(-bu^{-1})$.

To conclude, we would like to emphasize the close relationship between our results and those obtained for relevant extended perturbations [27, 30]. In these systems, the couplings deviate from the bulk ones by $AL^{-y}$, where $L$ is the distance to the surface. For the aperiodic systems, this corresponds to the average deviation per bond $\Delta(L)/L \sim L^{\omega-1}$ at a length scale $L$. Relevant perturbations are obtained for $y < 1$, which corresponds to $\omega > 0$. Close to the critical point, the short-distance structure of the perturbation is irrelevant and the surface singularities are the same for both systems.

\*\*\*

This work was supported by the CNRS and the Hungarian Academy of Sciences through an exchange program. FI was supported by the Hungarian National Research Fund under grant No. OTKA TO12830.REFERENCES

[1] SHECHTMAN D., BLECH I., GRATIAS D. and CAHN J. W., *Phys. Rev. Lett.*, **53** (1984) 1951.
[2] GELLERMANN W., KOHMOTO M., SUTHERLAND B. and TAYLOR P. C., *Phys. Rev. Lett.*, **72** (1994) 633.
[3] HENLEY C. L., *Comments Condens. Matt. Phys.*, **13** (1987) 59. For early work on random and arbitrary layered 2*d* Ising models see McCoy B. M. and Wu T. T., *Phys. Rev.*, **176** (1968) 631; AU-YANG H. and McCoy B. M., *Phys. Rev. B*, **10** (1974) 886.
[4] GODRÈCHE C., LUCK J. M. and ORLAND H. J., *J. Stat. Phys.*, **45** (1986) 777.
[5] OKABE Y. and NIIZEKI K., *J. Phys. Soc. Jpn.*, **57** (1988) 1536.
[6] SØRENSEN E. S., JARIĆ M. V. and RONCHETTI M., *Phys. Rev. B*, **44** (1991) 9271.
[7] SAKAMOTO S., YONEZAWA F., AOKI K., NOSÉ S. and HORI M., *J. Phys. A*, **22** (1989) L-705.
[8] ZHANG C. and DE'BELL K., *Phys. Rev. B*, **47** (1993) 8558.
[9] LANGIE G. and IGLÓI F., *J. Phys. A*, **25** (1992) L-487.
[10] IGLÓI F., *J. Phys. A*, **21** (1988) L-911.
[11] DORIA M. M. and SATIJA I. I., *Phys. Rev. Lett.*, **60** (1988) 444.
[12] BENZA G. V., *Europhys. Lett.*, **8** (1989) 321.
[13] HENKEL M. and PATKÓS A., *J. Phys. A*, **25** (1992) 5223.
[14] LIN Z. and TAO R., *J. Phys. A*, **25** (1992) 2483.
[15] TURBAN L. and BERCHE B., *Z. Phys. B*, **92** (1993) 307.
[16] HENLEY C. L. and LIPOWSKY R., *Phys. Rev. Lett.*, **59** (1987) 1679.
[17] GARG A. and LEVINE D., *Phys. Rev. Lett.*, **59** (1987) 1683.
[18] TRACY C. A., *J. Phys. A*, **21** (1988) L-603.
[19] HARRIS A. B., *J. Phys. C*, **7** (1974) 1671.
[20] LUCK J. M., *J. Stat. Phys.*, **72** (1993) 417.
[21] IGLÓI F., *J. Phys. A*, **26** (1993) L-703.
[22] LUCK J. M., *Europhys. Lett.*, **24** (1993) 359.
[23] DEKKING M., MENDÈS-FRANCE M. and VAN DER POORTEN A., *Math. Intelligencer*, **4** (1983) 130.
[24] TURBAN L., IGLÓI F. and BERCHE B., *Phys. Rev. B*, **49** (1994) 12695.
[25] KOGUT J., *Rev. Mod. Phys.*, **51** (1979) 659.
[26] LIEB E. H., SCHULTZ T. D. and MATTIS D. C., *Ann. Phys. (N. Y.)*, **16** (1961) 406.
[27] PESCHEL I., *Phys. Rev. B*, **30** (1984) 6783.
[28] PFEUTY P., *Phys. Lett. A*, **72** (1979) 245.
[29] DUMONT J. M., in *Number Theory and Physics, Spring. Proc. Physics*, Vol. **47**, edited by J. M. LUCK, P. MOUSSA and M. WALDSCHMIDT, (Springer, Berlin) 1990, p. 185.
[30] IGLÓI F., PESCHEL I. and TURBAN L., *Adv. Phys.*, **42** (1993) 683.